\documentclass[letterpaper, 10pt, conference]{ieeeconf}  

\newif\ifdraft
\draftfalse

\IEEEoverridecommandlockouts                             

\usepackage{graphics} 
\usepackage{epsfig} 
\usepackage{amsmath} 
\usepackage{amssymb}  
\usepackage{subcaption}  
\usepackage{lipsum}
\usepackage{ulem}
\usepackage{siunitx}
\usepackage{soul}
\usepackage[hidelinks]{hyperref}
\usepackage{textcomp}
\usepackage{mathtools}
\DeclarePairedDelimiterX{\norm}[1]{\lVert}{\rVert}{#1}
\usepackage[dvipsnames]{xcolor}
\usepackage{booktabs}  
\usepackage{multirow}
\usepackage{makecell}

\ifdraft
    \makeatletter
    \newcommand{\revised}[2][]{%
    \begingroup
    \sethlcolor{yellow}%
    \protected@edef\@tempa{#2}%
    \ifmmode
        \colorbox{yellow!30}{$\displaystyle #2$}%
    \else
        \hl{#2}%
    \fi
    \endgroup
    }
    \makeatother

\else
    \newcommand{\revised}[1]{#1}       
    \newcommand{\revmargin}[1]{}       

\fi

\DeclareMathAlphabet{\mathpzc}{OT1}{pzc}{m}{it}

\newcommand{\mbb}[1]{{\ensuremath{\mathbb{#1}}}}  
\newcommand{\mf}[1]{{\ensuremath{\mathfrak{#1}}}}  
\newcommand{\mc}[1]{{\ensuremath{\mathcal{#1}}}}  
\newcommand{\mr}[1]{{\ensuremath{\mathrm{#1}}}}   
\newcommand{\mb}[1]{{\ensuremath{\mathbf{#1}}}}   
\newcommand{\bs}[1]{\ensuremath{\boldsymbol{#1}}} 

\newcommand{\mring}[1]{\mathring{#1}}
\newcommand{\T}[0]{\mathsf{T}}

\usepackage{tikz}
\usetikzlibrary{shapes,arrows}
\usetikzlibrary{matrix} 
\newcommand{\suma}{\Large$+$}

\newcommand{\Rc}[0]{\mathbf{R}_{\mathcal{C}}}

\title{\LARGE \bf Equivariant Filter Cascade for Relative Attitude, Target's Angular Velocity, and Gyroscope Bias Estimation}

\author{Gil Serrano, Pedro Lourenço, Bruno J. Guerreiro, and Rita Cunha
\thanks{The work of Gil Serrano was supported by the PhD Grant from MIT Portugal and Funda\c{c}{\~a}o para a Ci{\^e}ncia e a Tecnologia (FCT), Portugal [DOI: 10.54499/PRT/BD/154275/2022]. This work was also supported by FCT, Portugal through LARSyS [DOI: 10.54499/LA/P/0083/2020] and project CAPTURE [DOI: 10.54499/PTDC/EEI-AUT/1732/2020]. (\textit{Corresponding author: Gil Serrano.})}
\thanks{Gil Serrano, Bruno J. Guerreiro and Rita Cunha are with the Institute for Systems and Robotics, Instituto Superior Técnico, Universidade de Lisboa, Lisboa, Portugal. (e-mail: \href{mailto:gil.serrano@tecnico.ulisboa.pt}{gil.serrano@tecnico.ulisboa.pt}; \href{mailto:bj.guerreiro@fct.unl.pt}{bj.guerreiro@fct.unl.pt}; \href{mailto:rita@isr.tecnico.ulisboa.pt}{rita@isr.tecnico.ulisboa.pt}).}
\thanks{Bruno J. Guerreiro is also with CTS/Uninova and LASI, School of Science and Technology, NOVA University Lisbon, Caparica, Portugal.}
\thanks{Pedro Lourenço is with the GNC Division, Flight Segment and Robotics, GMV, Lisboa, Portugal (e-mail: \href{mailto:palourenco@gmv.com}{palourenco@gmv.com}).}
}

\begin{document}

\tikzset{%
  block/.style    = {draw, thick, rectangle, minimum height = 3em,
    minimum width = 3em},
  sum/.style      = {draw, circle, node distance = 2cm}, 
  input/.style    = {coordinate}, 
  output/.style   = {coordinate} 
}

\maketitle
\thispagestyle{empty}
\pagestyle{empty}

\begin{abstract}

Rendezvous and docking between a chaser spacecraft and an uncooperative target, such as an inoperative satellite, require synchronization between the chaser spacecraft and the target. 
In these scenarios, the chaser must estimate the relative attitude and angular velocity of the target using onboard sensors, 
in the presence of gyroscope bias. 
In this work, we propose a cascade of Equivariant Filters (EqF) to address this problem. The first stage of the cascade estimates the chaser's attitude and the bias, using measurements from a star tracker, while the second stage of the cascade estimates the relative attitude and the target's angular velocity, using observations of two known, non-collinear vectors fixed in the target frame. 
The stability of the EqF cascade is theoretically analyzed and simulation results demonstrate the filter cascade's performance. 
\end{abstract}

\section{Introduction} \label{sec:introduction}

In recent years, there has been a significant increase in the number of satellites in orbit\;\cite{esa_space_debris_office_2025}. The accelerated rate of deployments worsens the problem of space debris in strategic orbits, which decreases the operational lifetime of the satellites\;\cite{ailor_effect_2017}. Therefore, there is a growing interest in On-Orbit Servicing and Active Debris Removal technologies to tackle this challenge\;\cite{hatty_viability_2022}.
In such missions, a chaser spacecraft will have to approach the target spacecraft and synchronize its motion before performing the planned procedures. Thus, the chaser needs to estimate the relative attitude and angular velocity of the target. To this effect, the chaser can use visual data acquired via conventional or event cameras, tracking features or fiducial markers on the target \;\cite{pesce_stereovision-based_2017, vela_pose_2022}, as well as inertial data obtained via an inertial measurement unit.

In the proposed scenario, when estimating the target's angular velocity in the presence of gyroscope bias, it is important to also estimate said bias; otherwise, the system states are not fully observable. If the bias is unknown and the gyroscope measurements of the chaser's angular velocity are used directly as filter inputs, then the target's angular velocity is unobservable and the estimates will be inaccurate. As such, many strategies that simultaneously estimate the states and the gyroscope bias have been proposed in the literature, with most being based on Kalman filtering\;\cite{crassidis_survey_2007}.
For example, considering the scenario of a rendezvous with a tumbling target spacecraft, Zhang et al.\;\cite{zhang_relative_2015} proposed an extended Kalman filter to estimate the relative pose and velocity, as well as the gyroscope bias of a chaser spacecraft, using line-of-sight measurements from an optical sensor.
In\;\cite{filipe_extended_journal_2015}, a dual quaternion multiplicative extended Kalman filter is presented for spacecraft pose, velocity, and accelerometer and gyroscope bias estimation, for uncooperative satellite proximity operations. This approach assumes that the chaser spacecraft has access to discrete-time measurements of the relative pose. By using the multiplicative method, the attitude estimate naturally conforms to the geometric constraints inherent to the problem.
In\;\cite{zivan_dual_2022}, an extended Kalman filter, also based on the dual quaternion representation, is designed for spacecraft relative pose and gyroscope bias estimation. Although the dual quaternion representation imposes geometric constraints on the filter state, these properties are enforced in an \textit{ad hoc} manner, rather than directly exploited by the proposed filter.

In our previous work\;\cite{serrano_eqf_relative_attitude_2025}, we developed an equivariant filter to estimate the relative attitude between two spacecraft, a target and a chaser, and the target's angular velocity, which was assumed constant in the target's reference frame. We also assumed that the chaser's angular velocity was known, e.g., through gyroscope readings, and that the target's body had physical features such that two non-collinear vectors could be measured, e.g., via a camera. In spite of accurately estimating the two desired states, the filter assumed an unbiased reading of the chaser's angular velocity. In this paper, we extend our approach by deriving another equivariant filter that estimates the gyroscope bias, as well as the chaser's attitude, using star tracker measurements. The two filters are then set up in a cascaded fashion, such that the bias estimate of the first filter is used to get unbiased gyroscope measurements of the chaser's angular velocity.
Equivariant filters that estimate input measurement biases along with system states such as orientation, position, or velocity, have been proposed in\;\cite{fornasier_overcoming_2022,fornasier_equivariant_2022,van_goor_eqvio_2023}. However, these approaches simultaneously estimate every quantity, whereas our method uses a multi-stage, cascaded structure. Not only does this simplify the derivation of the filter at each stage but it also allows for a more direct integration of measurements from different sensors, e.g., a camera and a star tracker. Moreover, while the symmetry groups in the aforementioned works are defined as semi-direct products of a given Lie group and its Lie algebra, we propose a simpler symmetry group.

The remainder of this paper is organized as follows: in Section\;\ref{sec:problem_statement}, we define the necessary reference frames, as well as the system and measurement models that will be used to design the filters; in Section\;\ref{sec:equivariant_filter_cascade}, we derive the equivariant filters and explain how they are connected in a cascaded manner; in Section\;\ref{sec:stability_analysis}, we analyze the stability of the full design; in Section\;\ref{sec:simulation_results} we present simulation results, in which low-rate measurements were considered; finally, in Section\;\ref{sec:conclusions}, we provide closing remarks.

\section{Problem Statement}\label{sec:problem_statement}

In this section, the necessary reference frames are defined and both the system model and the measurement model are presented.

\subsection{Reference Frames and Attitude Kinematics} 
\label{subsec:reference_frames_and_attitude_kinematics}
We consider two rotating body-fixed frames, the target's frame $\{\mc{T}\}$ and the chaser's frame $\{\mc{C}\}$, as well as an inertial frame $\{\mc{I}\}$. The attitudes of the target and chaser's frames with respect to the inertial frame are denoted by the rotation matrices ${}^\mc{I}_\mc{T}\mb{R}$ and ${}^\mc{I}_\mc{C}\mb{R}$, respectively. The rotation matrix ${{}^\mc{T}_\mc{C}\mb{R}}$ describes the attitude of the chaser with respect to the target. 

The angular velocities of the target and chaser, with respect to the inertial frame, expressed in their respective frames are represented by ${\bs{\omega}_\mc{T}}$ and ${\bs{\omega}_\mc{C}}$. We define ${\bs{\omega} \equiv {}^\mc{C}_\mc{T}\mb{R}\,\bs{\omega}_\mc{T}}$, which is the angular velocity of the target, with respect to the inertial frame, expressed in the chaser's frame. In this work, we consider the scenario where the target angular velocity is unknown but constant, i.e., $\dot{\bs{\omega}}_\mc{T} = \mb{0}$, consistent with a body rotating about the principal axis with the largest moment of inertia.

The attitude kinematics for the aforementioned frames are given by
\begin{align}
{}^\mc{I}_\mc{T}\dot{\mb{R}} \,&=\, {}^\mc{I}_\mc{T}\mb{R}\,{\bigl(\bs{\omega}_\mc{T}\bigr)}^{\wedge}\,,\\
{}^\mc{I}_\mc{C}\dot{\mb{R}} \,&=\, {}^\mc{I}_\mc{C}\mb{R}\,{(\bs{\omega}_\mc{C}\bigr)}^{\wedge}\,,\\
{}^\mc{T}_\mc{C}\dot{\mb{R}} \,&=\, {}^\mc{T}_\mc{C}\mb{R}\,{\bigl(\bs{\omega}_{\mc{C}} - \bs{\omega}\bigr)}^\wedge\,, 
\end{align}
where $(\mb{x})^{\wedge}$, for ${\mb{x}\in\mbb{R}^{3}}$, is the skew-symmetric matrix constructed from the vector $\mb{x}$, such that ${(\mb{x})^{\wedge}\mb{y} = \mb{x}\times\mb{y}}$, ${\forall\mb{y}\in\mbb{R}^3}$. The inverse operation to ${(\cdot)^{\wedge}}$ is denoted as ${(\cdot)^{\vee}}$, such that ${{((\mb{x})^{\wedge})}^{\vee} = \mb{x}}$, ${\forall\mb{x}\in\mbb{R}^{3}}$.

\subsection{System Model} 
\label{subsec:system_model}

To ease notation, we define ${\Rc\equiv{}^\mc{I}_\mc{C}\mb{R}}$ and ${\mb{R}\equiv{}^\mc{T}_\mc{C}\mb{R}}$. Let the chaser's angular velocity be denoted by ${\mb{u}\equiv\bs{\omega}_{\mc{C}}}$ and the measurement of this quantity by a gyroscope be represented by ${\bar{\mb{u}}}$, given by
\begin{equation}
\bar{\mb{u}} = \mb{u} + \mb{b} 
\,,
\end{equation}
where ${\mb{b}}$ is a constant bias. 

Since we are going to develop a cascade of equivariant filters, we divide the system model into two parts: the first concerns the chaser's attitude, ${\mb{R}_{\mc{C}}}$, and the gyroscope bias, $\mb{b}$, while the second focuses on the relative attitude, ${\mb{R}}$, and the target's angular velocity, ${\bs{\omega}}$.

For the first part of the system model, we formulate: a state ${\bigl(\Rc,\,\mb{b}\bigr) \in \mc{M}^{(1)}  = \mc{SO}(3) \times \mbb{R}^3}$, where ${\mc{SO}(3)}$ is the $\mr{SO}(3)$-torsor\;\cite{mahony_observers_2013} and $\mr{SO}(3)$ is the special orthogonal group; and an input ${\bar{\mb{u}} \in \mbb{L}^{(1)} = \mbb{R}^3}$, where $\mc{M}^{(1)}$ and $\mbb{L}^{(1)}$ are the state and input manifolds, respectively. Thus, the first part of the system under consideration is ${(\dot{\mb{R}}_{\mc{C}}, \dot{\mb{b}}) = f^{(1)}((\Rc,\mb{b}), \bar{\mb{u}})}$, as described by
\begin{align}
    \dot{\mb{R}}_{\mc{C}}\,&=\,
    \mb{R}_{\mc{C}}\,\bigl(\bar{\mb{u}} - \mb{b} 
    \bigr)^{\wedge}\,,\label{eq:system_model_Rc}\\
    \dot{\mb{b}} \,&=\, \mb{0}\,.\label{eq:system_model_bias}
\end{align}

Similarly, for the second part of the system model, we formulate a state ${\bigl(\mb{R},\,\bs{\omega}\bigr) \in \mc{M}^{(2)}  = \mc{SO}(3) \times \mbb{R}^3}$ and input ${\mb{u} \in \mbb{L}^{(2)} = \mbb{R}^3}$. As such, the second part of the system under consideration is ${(\dot{\mb{R}}, \dot{\bs{\omega}}) = f^{(2)}((\mb{R},\bs{\omega}), \mb{u})}$ and is given by
\begin{align}
    \dot{\mb{R}} \,&=\, \mb{R}\,\bigl(\mb{u} - \bs{\omega} \bigr)^\wedge\,, \label{eq:system_model_R} \\
    \dot{\bs{\omega}} \,&=\, (\bs{\omega})^\wedge\mb{u}\,.\label{eq:system_model_omega}
\end{align}
The superscripts ${}^{(1)}$ and ${}^{(2)}$ are used throughout this work to refer to the first and second stage filters, respectively.
\subsection{Measurement Model} 
\label{subsec:measurement_model}
We also define two measurement models, which will be used separately in each stage of the filter cascade.

For the first stage of the cascade, we assume that the chaser is equipped with a star tracker, which is capable of giving measurements of the chaser's attitude, ${\Rc}$. However, in the equivariant filter formulation, the output space, ${\mc{Y}^{(1)}}$, is required to be embedded in ${\mbb{R}^{n}}$. Hence, we manipulate the measurement, making use of the canonical basis vectors for ${\mbb{R}^3}$, ${\{\mb{e}_1,\mb{e}_2,\mb{e}_3\}}$, and define the output ${\mb{y}^{(1)}=h^{(1)}\bigl(\Rc,\mb{b}\bigr)}$, ${h^{(1)}:\mc{M}^{(1)}\to\mc{Y}^{(1)}}$ and ${\mc{Y}^{(1)} = \mbb{R}^3\times\mbb{R}^3\times\mbb{R}^3}$, as being
\begin{equation}
    \mb{y}^{(1)} = h^{(1)}\bigl(\Rc, \mb{b}\bigr) = \bigl(\Rc^\T\,\mb{e}_1,\,\Rc^\T\,\mb{e}_2,\,\Rc^\T\,\mb{e}_3\bigr).
    \label{eq:system_output_y_1}
\end{equation}

For the second stage of the filter cascade, we assume that the chaser is able to measure two non-collinear reference vectors on the target's body, denoted by $\mb{d}_1$ and $\mb{d}_2$, when expressed in the chaser's frame, $\{\mc{C}\}$. These vectors are constant when expressed in the target's frame, $\{\mc{T}\}$, and are denoted by $\mring{\mb{d}}_1$, $\mring{\mb{d}}_2\in\mbb{R}^3$. Thus, the output of the second part of the system, represented by ${\mb{y}^{(2)} = h^{(2)}\bigl(\mb{R}, \bs{\omega}\bigr)}$, where $h^{(2)}:\mc{M}^{(2)}\to\mc{Y}^{(2)}$ and ${\mc{Y}^{(2)}=\mbb{R}^3\times\mbb{R}^3}$, is given by
\begin{equation}
    \mb{y}^{(2)} = h^{(2)}\bigl(\mb{R}, \bs{\omega}\bigr) = \bigl(\mb{d}_1,\, \mb{d}_2\bigr) = \bigl(\mb{R}^\T\,\mring{\mb{d}}_1,\mb{R}^\T\,\mring{\mb{d}}_2\bigr).
    \label{eq:system_output_y_2}
\end{equation}







\section{Equivariant Filter Cascade}\label{sec:equivariant_filter_cascade}

We derive the two equivariant filters which will be used in the cascade. The first stage filter estimates the chaser's attitude and the gyroscope bias, while the the second stage filter estimates the relative attitude and the target's angular velocity. The bias estimate of the first stage filter is subtracted from the gyroscope data, in order to feed an unbiased angular velocity measurement to the second stage filter. We also analyze the conditions for the convergence of the cascade.

\subsection{First Stage Equivariant Filter: Chaser's Attitude and Gyroscope Bias Estimation} \label{subsec:first_stage_eqf}

The symmetry group used in this problem is ${\mr{G} = \mr{SE}(3)}$, the special Euclidean group, with Lie-algebra ${\mf{g} = \mf{se}(3)}$, and ${(\mb{A}, \mb{a})\in \mr{G}}$ is the first stage equivariant filter's group state.
We also define the origin ${(\mring{\mb{R}}_{\mc{C}}, \mring{\mb{b}}) := (\mb{I}, \mb{0})}$ for the coordinate system on the state manifold $\mc{M}^{(1)}$, which is needed to map the filter's estimates from the group to the manifold.

\subsubsection{State, Input, and Output Actions}

To derive the equivariant filter, we first need to define a right state action ${\phi^{(1)}:\mr{G}\times\mc{M}^{(1)}\to\mc{M}^{(1)}}$, a right input action ${\psi^{(1)}:\mr{G}\times\mbb{L}^{(1)}\to\mbb{L}^{(1)}}$, and a right output action ${\rho^{(1)}:\mr{G}\times\mc{Y}^{(1)}\to\mc{Y}^{(1)}}$. These maps are given by
\begin{equation}
\begin{split}
    \phi^{(1)}\big((\mb{A}, \mb{a}),(\Rc,\mb{b})\big) &= \big(\Rc\mb{A},\mb{A}^\T(\mb{b} - \mb{a})\big),\\
    \psi^{(1)}\big((\mb{A}, \mb{a}),\bar{\mb{u}}\big) &= \mb{A}^\T(\bar{\mb{u}} - \mb{a}),\\
    \rho^{(1)}\big((\mb{A}, \mb{a}),\mb{y}^{(1)}\big) &= \mb{A}^\T \mb{y}^{(1)}.
\end{split}
    \label{eq:eqf1_right_actions}
\end{equation}
The manifold state can be recovered from the group state using the state action and the coordinate system origin as ${(\Rc,\mb{b}) = \phi^{(1)}_{(\mring{\mb{R}}_{\mc{C}},\mring{\mb{b}})}(\mb{A}, \mb{a}) = \phi^{(1)}\big((\mb{A}, \mb{a}),(\mring{\mb{R}}_{\mc{C}},\mring{\mb{b}})\big)}$.

\subsubsection{Equivariant System Lift}

The equivariant filter is defined on the group $\mr{G}$ with group state estimates ${(\hat{\mb{A}},\hat{\mb{a}})\in \mr{G}}$. The lifted system dynamics on the group are given by
\begin{equation}
\begin{split}
    \dot{\hat{\mb{A}}} &= \hat{\mb{A}}\, \Lambda_{\mb{A}}\big(\phi^{(1)}_{(\mring{\mb{R}}_{\mc{C}},\mring{\mb{b}})}(\hat{\mb{A}}, \hat{\mb{a}}), \bar{\mb{u}}\big) + \Delta_\mb{A} \hat{\mb{A}},\\
    \dot{\hat{\mb{a}}} &= \hat{\mb{A}}\,\Lambda_{\mb{a}}\big(\phi^{(1)}_{(\mring{\mb{R}}_{\mc{C}},\mring{\mb{b}})}(\hat{\mb{A}}, \hat{\mb{a}}), \bar{\mb{u}}\big) + \Delta_\mb{A}\hat{\mb{a}} + \delta_\mb{a},
    \label{eq:eqf1_group_dynamics}
\end{split}
\end{equation}
where ${(\Lambda_{\mb{A}}, \Lambda_{\mb{a}})=\Lambda^{(1)}}$ is the equivariant system lift and ${(\Delta_{\mb{A}},\delta_{\mb{a}})=\Delta^{(1)}}$ are the corrections terms to be derived. The lift ${\Lambda^{(1)}=(\Lambda_{\mb{A}}, \Lambda_{\mb{a}}):\mc{M}^{(1)}\times\mbb{L}^{(1)}\to\mf{g}}$ is given by
\begin{align}
     \Lambda_{\mb{A}}((\mb{R}_{\mc{C}},\mb{b}),\bar{\mb{u}}) &= (\bar{\mb{u}} - \mb{b})^{\wedge} , \\
    \Lambda_{\mb{a}}((\mb{R}_{\mc{C}},\mb{b}),\bar{\mb{u}}) &= -(\bar{\mb{u}})^{\wedge} \mb{b}.
    \label{eq:eqf1_lift}
\end{align}

The equivariant filter dynamics also include an equation related to a Ricatti state, $\bs{\Sigma}^{(1)}\in\mbb{S}^{6}_{+}$, which is a symmetric, positive definite matrix. The Riccati state dynamics, with initial value ${\bs{\Sigma}^{(1)}(0) = \bs{\Sigma}^{(1)}_{0}\in\mbb{S}^{6}_{+}}$, are
\begin{equation}
\begin{split}
    &\dot{\bs{\Sigma}}^{(1)} =  \mring{\mb{A}}^{(1)}_t \bs{\Sigma}^{(1)} + \bs{\Sigma}^{(1)} {{\mring{\mb{A}}}^{(1)^\T}}_t + \mb{M}^{(1)}_t \\
    &\hspace{25mm}- \bs{\Sigma}^{(1)} {{\mb{C}_t}}^{(1)^\T} {\mb{N}^{(1)^{-1}}_t} {{\mb{C}_t}^{(1)}} \bs{\Sigma}^{(1)},
\end{split}
\label{eq:eqf1_riccati_state_equation}
\end{equation}
where ${\mb{M}^{(1)}_t\in\mbb{S}^{6}_{+}}$ and ${\mb{N}^{(1)}_t\in\mbb{S}^{9}_{+}}$ are the state and output gain matrices, respectively, and the matrices $\mring{\mb{A}}^{(1)}_t$ and ${{\mb{C}_t}^{(1)}}$ will be defined next.

\subsubsection{Error System and Correction Terms}
To develop the filter, we must define three types of system error: i) the error on the group, ${\mb{E}^{(1)}=(\mb{E}_{\mb{A}},\mb{E}_{\mb{a}})}$; ii) the error on the manifold, ${\mb{e}^{(1)} =(\mb{e}_{\Rc}, \mb{e}_{\mb{b}})}$; and iii) the error on the tangent space at the previously defined origin, ${\bs{\varepsilon}^{(1)} =(\bs{\varepsilon}_{\Rc},\bs{\varepsilon}_{\mb{b}})}$.
Let us first define the group error, which is
\begin{equation}
\begin{split}
    \mb{E}^{(1)} &= (\mb{A},\mb{a})\cdot(\hat{\mb{A}},\hat{\mb{a}})^{-1} = (\mb{A}\hat{\mb{A}}^\T, - \mb{A}\hat{\mb{A}}^\T \hat{\mb{a}} + \mb{a} )\,.
\end{split}
\label{eq:eqf1_group_error}
\end{equation}
Then, the error on the manifold is obtained via the state action $\phi^{(1)}$, acting on the origin, according to 
\begin{equation}
    \begin{split}
        \mb{e}^{(1)} = \phi_{(\mring{\mb{R}}_\mc{C}, \mring{\mb{b}})}(\mb{E}_{\mb{A}},\mb{E}_{\mb{a}})\,.
    \end{split}
\label{eq:eqf1_manifold_state_error}
\end{equation}
Lastly, to define an error on the tangent space at the origin, we fix a local coordinate chart ${\vartheta^{(1)}:\mc{N}^{(1)}_{(\mring{\mb{R}}_{\mc{C}}, \mring{\mb{b}})} \to \mc{T}_{(\mring{\mb{R}}_{\mc{C}}, \mring{\mb{b}})}\mc{M}^{(1)}}$, where ${\mc{N}^{(1)}_{(\mring{\mb{R}}_{\mc{C}}, \mring{\mb{b}})}\subset\mc{M}^{(1)}}$ is a neighborhood of the origin $(\mring{\mb{R}}_{\mc{C}}, \mring{\mb{b}})$ and ${\mc{T}_{(\mring{\mb{R}}_{\mc{C}}, \mring{\mb{b}})}\mc{M}^{(1)}\equiv \mf{g}}$ is the tangent space at the origin. Since $\mbb{R}^3\times\mbb{R}^3$ is isomorphic to $\mf{g}$, we adapt the coordinate chart such that ${\vartheta^{(1)}:\mc{N}^{(1)}_{(\mring{\mb{R}}_{\mc{C}}, \mring{\mb{b}})} \to \mbb{R}^3\times\mbb{R}^3}$. This allows us to conduct the remaining derivations in vector form. Thus, the local coordinates of the state error, ${\bs{\varepsilon}^{(1)} = (\bs{\varepsilon}_{\Rc},\bs{\varepsilon}_{\mb{b}}) \in \mbb{R}^3 \times \mbb{R}^3}$, are given by
\begin{equation}
    \bs{\varepsilon}^{(1)} = \vartheta^{(1)}(\mb{e}^{(1)}) = \big(\text{log}(\mb{e}_{\Rc})^{\vee}, {\mb{e}_{\mb{b}}}\big)\,.
    \label{eq:eqf1_local_coordinates_error}
\end{equation}

The dynamics of the error on the tangent space at the origin are described by
\begin{equation}
\begin{split}
    &\dot{\bs{\varepsilon}}^{(1)} = D\vartheta^{(1)} \cdot D\phi^{(1)}_{\mb{e}^{(1)}}\Big(\Lambda^{(1)}\big(\mb{e}^{(1)},\mring{\bar{\mb{u}}}\big) \\
    &\hspace{35mm}- \Lambda^{(1)}\big((\mring{\mb{R}_{\mc{C}}},\mring{\mb{b}}),\mring{\bar{\mb{u}}}\big)\Big) \\
    &\hspace{35mm}- D\vartheta^{(1)} \cdot D\phi^{(1)}_{\mb{e}^{(1)}}\big(\Delta^{(1)}\big),
\end{split}
\label{eq:eqf1_local_coordinates_error_dynamics}
\end{equation}
where $\mring{\bar{\mb{u}}} = \psi^{(1)}_{(\hat{\mb{A}},\hat{\mb{a}})^{-1}}(\bar{\mb{u}})$. We linearize the pre-observer part this equation around ${{\bs{\varepsilon}}^{(1)} = \mb{0}}$ and get 
\begin{equation}
    \dot{\bs{\varepsilon}}^{(1)} \approx \mring{\mb{A}}^{(1)}_t\bs{\varepsilon}^{(1)} - D\vartheta^{(1)} \cdot D\phi^{(1)}_{\mb{e}^{(1)}}\big(\Delta^{(1)}\big)\,,\\
\label{eq:eqf1_linearized_error_dynamics}    
\end{equation}
where the matrix ${\mring{\mb{A}}^{(1)}_t}$ is 
\begin{equation}
    \mring{\mb{A}}^{(1)}_t = \begin{bmatrix}
        \mb{0} & -\mb{I}\\
        \mb{0} & \big(\hat{\mb{A}} \bar{\mb{u}} + \hat{\mb{a}}\big)^{\wedge}
    \end{bmatrix}.
\end{equation}

Next, we define the output residual ${\Tilde{\mb{y}}^{(1)} = \mb{y}^{(1)}\big(\bs{\varepsilon}^{(1)}\big) - \hat{\mb{y}}^{(1)}}$. Linearizing ${\Tilde{\mb{y}}^{(1)}}$ around ${\bs{\varepsilon}^{(1)} = \mb{0}}$, gives
\begin{equation}
    \Tilde{\mb{y}}^{(1)} \approx {\mb{C}_t}^{(1)} \bs{\varepsilon}^{(1)}\,,
\end{equation}
where the matrix ${{\mb{C}_t}^{(1)}}$ is
\begin{equation}
    {\mb{C}_t}^{(1)} = \frac{1}{2} 
    \begin{bmatrix}
    \big(\mb{y}^{(1)}_{1} + \hat{\mb{y}}^{(1)}_{1}\big)^{\wedge} \hat{\mb{A}}^\T & \mb{0}\\
    \big(\mb{y}^{(1)}_{2} + \hat{\mb{y}}^{(1)}_{2}\big)^{\wedge} \hat{\mb{A}}^\T& \mb{0}\\
    \big(\mb{y}^{(1)}_{3} + \hat{\mb{y}}^{(1)}_{3}\big)^{\wedge} \hat{\mb{A}}^\T& \mb{0}\\
    \end{bmatrix}.
\end{equation}

Finally, the correction terms ${\Delta^{(1)}=(\Delta_\mb{A}, \delta_\mb{a})}$ are calculated according to
\begin{equation}
\begin{split}
    &(\Delta_{\mb{A}},\delta_{\mb{a}}) = {D\phi^{(1)}_{\mb{e}^{(1)}}}^\dagger \cdot\\
    & \hspace{12mm} D{\vartheta^{(1)}}^{-1} \big[\bs{\Sigma}^{(1)} {{\mb{C}_t}}^{(1)^\T} {\mb{N}^{(1)^{-1}}_t}\big(\mb{y}^{(1)}-\hat{\mb{y}}^{(1)}\big)\big],
\end{split}
\label{eq:eqf1_correction_term_delta}
\end{equation}
where ${{(\cdot)}^\dagger}$ designates a right inverse.

\subsection{Second Stage Equivariant Filter: Relative Attitude and Target's Angular Velocity Estimation} \label{subsec:second_stage_eqf}

For the second stage of the EqF cascade, we will use the filter derived in\;\cite{serrano_eqf_relative_attitude_2025}, which follows the same steps, \textit{mutatis mutandis}, as the first stage equivariant filter. As the derivation of the second stage filter is similar to the first stage filter, we will present a brief overview of the filter.

The symmetry group for the second filter is also $\mr{G}$, the origin for the coordinate system on the state manifold ${\mc{M}^{(2)}}$ is ${(\mring{\mb{R}}, \mring{\bs{\omega}}) = (\mb{I}, \mb{0})}$, and the group state is ${(\mb{Q},\mb{q})\in \mr{G}}$. 

\subsubsection{State, Input, and Output Actions}

To prove equivariance of the second part of the system, ${(\dot{\mb{R}}, \dot{\bs{\omega}}) = f^{(2)}((\mb{R},\bs{\omega}), \mb{u})}$, two virtual inputs have to be considered, ${\mb{v},\mb{w}\in\mbb{R}^3}$, which can be assumed equal to zero, in order to recover the original system. Ergo, the output manifold is ${\mbb{L}^{(2)} = \mbb{R}^3\times\mbb{R}^3\times\mbb{R}^3}$. More details can be found in\;\cite{serrano_eqf_relative_attitude_2025}.

The right state action ${\phi^{(2)}:\mr{G}\times\mc{M}^{(2)}\to\mc{M}^{(2)}}$, which allows us to take the filter estimates from the group, the right input action ${\psi^{(2)}:\mr{G}\times\mbb{L}^{(2)}\to\mbb{L}^{(2)}}$, and the right output action ${\rho^{(2)}:\mr{G}\times\mc{Y}^{(2)}\to\mc{Y}^{(2)}}$, are
\begin{equation}
\begin{split}
    \phi^{(2)}\big((\mb{Q}, \mb{q}),(\mb{R},\bs{\omega})\big) &= \big(\mb{R}\mb{Q},\mb{Q}^\T(\bs{\omega} - \mb{q})\big),\\
    \psi^{(2)}\big((\mb{Q}, \mb{q}),(\mb{u},\mb{v},\mb{w})\big) &= \big(\mb{Q}^\T\mb{u},\mb{Q}^\T(\mb{v}-\mb{q}),\\
    & \hspace{22mm} \mb{Q}^\T(\mb{w}+\mb{q})\big),\\
    \rho^{(2)}\big((\mb{Q}, \mb{q}),\mb{y}^{(2)}\big) &= \mb{Q}^\T \mb{y}^{(2)}.
\end{split}
\label{eq:eqf2_right_actions}
\end{equation}

\subsubsection{Equivariant System Lift}
The lifted equivariant filter dynamics, with group state ${(\hat{\mb{Q}}, \hat{\mb{q}})\in \mr{G}}$ and Riccati state ${\bs{\Sigma}^{(2)}\in\mbb{S}^{6}_{+}}$, are given by
\begin{equation}
\begin{split}
    \dot{\hat{\mb{Q}}} &= \hat{\mb{Q}}\, \Lambda_{\mb{Q}}\big(\phi_{(\mring{\mb{R}},\mring{\bs{\omega}})}(\hat{\mb{Q}}, \hat{\mb{q}}), (\mb{u},\mb{v},\mb{w})\big) + \Delta_\mb{Q} \hat{\mb{Q}},\\
    \dot{\hat{\mb{q}}} &= \hat{\mb{Q}}\,\Lambda_{\mb{q}}\big(\phi_{(\mring{\mb{R}},\mring{\bs{\omega}})}(\hat{\mb{Q}}, \hat{\mb{q}}), (\mb{u},\mb{v},\mb{w})\big) + \Delta_\mb{Q}\hat{\mb{q}} + \delta_\mb{q},\\
    \hspace{-1mm}\dot{\bs{\Sigma}}^{(2)}&=  {\mring{\mb{A}}_t}^{(2)} \bs{\Sigma}^{(2)} + \bs{\Sigma}^{(2)} {\mring{\mb{A}}_t}^{(2)^\T} + \mb{M}^{(2)}_t \\
    &\hspace{15mm}-\bs{\Sigma}^{(2)} {\mb{C}_t}^{(2)^\T}{\mb{N}^{(2)^{-1}}_t} {\mb{C}_t}^{(2)} \bs{\Sigma}^{(2)},
\end{split}
\label{eq:eqf2_group_dynamics}
\end{equation}
where ${(\Lambda_{\mb{Q}}, \Lambda_{\mb{q}}) = \Lambda^{(2)}}$ is the the equivariant lift, ${(\Delta_{\mb{Q}},\delta_{\mb{q}}) = \Delta^{(2)}}$ are the correction terms, ${\mb{M}^{(2)}_t\in\mbb{S}^{6}_{+}}$ and ${\mb{N}^{(2)}_t\in\mbb{S}^{6}_{+}}$ are the state and output gain matrices, respectively, and $\mring{\mb{A}}_t^{(2)}$ and ${\mb{C}_t}^{(2)}$ will be defined next. The lift ${\Lambda^{(2)}=(\Lambda_{\mb{Q}}, \Lambda_{\mb{q}}):\mc{M}^{(2)}\times\mbb{L}^{(2)}\to\mf{g}}$ is given by
\begin{align}
    \Lambda_{\mb{Q}}((\mb{R},\bs{\omega}),(\mb{u},\mb{v},\mb{w})) &= (\mb{u} - \bs{\omega} + \mb{v})^{\wedge}, \\
    \Lambda_{\mb{q}}((\mb{R},\bs{\omega}),(\mb{u},\mb{v},\mb{w})) &= (\mb{u})^\wedge\mb{w} + (\bs{\omega})^\wedge\mb{v}.
    \label{eq:eqf2_equivariant_lift}
\end{align}

\subsubsection{Error System and Correction Terms}

As in the first stage filter, we require the definition of the group error, ${\mb{E}^{(2)} = (\mb{E}_{\mb{Q}},\mb{E}_{\mb{q}})}$, the manifold error, ${\mb{e}^{(2)} = (\mb{e}_{\mb{R}},\mb{e}_{\bs{\omega}})}$, the local coordinates error, ${\bs{\varepsilon}^{(2)} = (\bs{\varepsilon}_{\mb{R}}, \bs{\varepsilon}_{\bs{\omega}})}$, and the map ${\vartheta^{(2)}}$\;\cite{serrano_eqf_relative_attitude_2025}. 
The dynamics of the error on the tangent space at the origin are described by
\begin{equation}
\begin{split}
    &\dot{\bs{\varepsilon}}^{(2)} = D\vartheta^{(2)} \cdot D\phi^{(2)}_{\mb{e}^{(2)}}\Big(\Lambda^{(2)}\big(\mb{e}^{(2)},(\mring{\mb{u}},\mring{\mb{v}},\mring{\mb{w}})\big) \\
    &\hspace{35mm}- \Lambda^{(2)}\big((\mring{\mb{R}},\mring{\bs{\omega}}),(\mring{\mb{u}},\mring{\mb{v}},\mring{\mb{w}})\big)\Big) \\
    &\hspace{35mm}- D\vartheta^{(2)} \cdot D\phi^{(2)}_{\mb{e}^{(2)}}\big(\Delta^{(2)}\big).
\end{split}
\label{eq:eqf2_local_coordinates_error_dynamics}
\end{equation}
where ${(\mring{\mb{u}},\mring{\mb{v}},\mring{\mb{w}}) = \psi_{(\hat{\mb{Q}},\hat{\mb{q}})^{-1}}(\mb{u},\mb{v},\mb{w})}$.
We linearize the local coordinates error pre-observer dynamics and output residual around ${\bs{\varepsilon}}^{(2)} = \mb{0}$, and get the linearized dynamics
\begin{equation}
\begin{split}
    \dot{\bs{\varepsilon}}^{(2)} &\approx \mring{\mb{A}}^{(2)}_t\bs{\varepsilon}^{(2)} - D\vartheta^{(2)} \cdot D\phi^{(2)}_{\mb{e}^{(2)}}\big(\Delta^{(2)}\big),\\
    \Tilde{\mb{y}}^{(2)} &\approx {\mb{C}_t}^{(2)} \bs{\varepsilon}^{(2)}\,,
\end{split}
\label{eq:eqf2_linearized_error_dynamics}
\end{equation}
where the matrices ${\mring{\mb{A}}^{(2)}_t}$ and ${{\mb{C}_t}^{(2)}}$ are
\begin{equation}
\begin{split}
    \mring{\mb{A}}_t^{(2)} &= \begin{bmatrix}
        \mb{0} & -\mb{I}\\
        \mb{0} & (\hat{\mb{q}})^{\wedge}
    \end{bmatrix}\,,\\
    {\mb{C}_t}^{(2)} &= \frac{1}{2} 
    \begin{bmatrix}
    \big(\mb{y}^{(2)}_1 + \hat{\mb{y}}^{(2)}_1\big)^{\wedge} \hat{\mb{Q}}^\T & \mb{0}\\
    \big(\mb{y}^{(2)}_2 + \hat{\mb{y}}^{(2)}_2\big)^{\wedge} \hat{\mb{Q}}^\T& \mb{0}\\
    \end{bmatrix}\,.
\end{split}
\label{eq:eqf2_linearized_matrices}
\end{equation}
The correction terms ${\Delta^{(2)}=(\Delta_\mb{Q}, \delta_\mb{q})}$ are calculated by
\begin{equation}
\begin{split}
    &(\Delta_{\mb{Q}},\delta_{\mb{q}}) = {D\phi^{(2)}_{\mb{e}^{(2)}}}^\dagger \cdot\\
    & \hspace{12mm} D{\vartheta^{(2)}}^{-1} \big[\bs{\Sigma}^{(2)} {{\mb{C}_t}}^{(2)^\T} {\mb{N}^{(2)^{-1}}_t}\big(\mb{y}^{(2)}-\hat{\mb{y}}^{(2)}\big)\big]\,.
\end{split}
\label{eq:eqf2_correction_term_delta}
\end{equation}

\section{Stability Analysis of the Filter Cascade} \label{sec:stability_analysis}

\begin{figure}
    \centering
    \resizebox{\columnwidth}{!}{%
        \begin{tikzpicture}[auto, thick, node distance=1.5cm, >=triangle 45]
\draw [color=BrickRed, fill=BrickRed!10,,thick](3,1) rectangle (12,-2.5);
\node at (3,1) [above=2mm, right=0mm] {\large \textsc{\textcolor{BrickRed}{\textbf{first stage equivariant filter}}}};
\draw [color=ForestGreen, fill=ForestGreen!10, thick](3,-4) rectangle (12,-7.5);
\node at (3,-4) [above=2mm, right=0mm] {\large \textsc{\textcolor{ForestGreen}{\textbf{second stage equivariant filter}}}};
\node at (-1,1) {\large \textsc{\textcolor{RoyalBlue}{\textbf{sensors}}}};

\draw
	node at (-1,0)[block, color=RoyalBlue, fill=RoyalBlue!10, minimum width=28mm, name=startracker]{\Large \textcolor{black}{Star Tracker}}
    node at (-1,-1.5)[block, color=RoyalBlue, fill=RoyalBlue!10, minimum width=28mm] (gyro){\Large \textcolor{black}{Gyroscope}}
    node at (6,-0.75) [block, fill=white, minimum height=7em,minimum width=10em] (EqF1) {\Large $\begin{matrix}
        \dot{\hat{\mb{A}}} = \hat{\mb{A}}\Lambda_{\mb{A}} + \Delta_{\mb{A}}\hat{\mb{A}}\\
        \dot{\hat{\mb{a}}} = \hat{\mb{A}}\Lambda_{\mb{a}} + \Delta_{\mb{A}}\hat{\mb{a}} + \delta_{\mb{a}}
    \end{matrix}$}
    node at (3.6,0) [input, name=EqF1in1] {}
    node at (3.6,-1.5) [input, name=EqF1in2] {}
    node at (8.4,0) [output, name=EqF1out1] {}
    node at (8.4,-1.5) [output, name=EqF1out2] {}
    
    node at (11,-0.75) [block, fill=white, minimum height = 2cm] (phi1) {\Large $\phi^{(1)}_{(\mring{\mb{R}}_{\mc{C}}, \mring{\mb{b}})}$}
    node at (10.2,0) [input, name=phi1in1] {}
    node at (10.2,-1.5) [input, name=phi1in2] {}
    node at (11.8,0) [input, name=phi1out1] {}
    node at (11.8,-1.5) [input, name=phi1out2] {}

    node at (14,0) [input, name=stage1out1] {}
    node at (14,-1.5) [input, name=stage1out2] {}
;
    \draw[->](startracker) -- node [xshift=-4mm] {\Large $\Rc^\T\mb{e}_i$}(EqF1in1);
    \draw[->](gyro) -- node [xshift=-4mm] {\Large $\bar{\mb{u}}$} (EqF1in2);
    \draw[->](EqF1out1) -- node {\Large $\hat{\mb{A}}$} (phi1in1);
    \draw[->](EqF1out2) -- node {\Large $\hat{\mb{a}}$} (phi1in2);
    \draw[->](phi1out1) -- node {\Large $\hat{\mb{R}}_{\mc{C}}$} (stage1out1);
    \draw[->](phi1out2) -- node {\Large $\hat{\mb{b}}$} (stage1out2);

\draw
    node at (-1,-6.5)[block, color=RoyalBlue, fill=RoyalBlue!10, minimum width=28mm, name=camera]{\Large \textcolor{black}{Camera}}
    node at (6,-5.75) [block, fill=white, minimum height=7em,minimum width=10em] (EqF1) {\Large $\begin{matrix}
        \dot{\hat{\mb{Q}}} = \hat{\mb{Q}}\Lambda_{\mb{Q}} + \Delta_{\mb{Q}}\hat{\mb{Q}}\\
        \dot{\hat{\mb{q}}} = \hat{\mb{Q}}\Lambda_{\mb{q}} + \Delta_{\mb{Q}}\hat{\mb{q}} + \delta_{\mb{q}}
    \end{matrix}$}
    node at (3.55,-5) [input, name=EqF2in1] {}
    node at (3.55,-6.5) [input, name=EqF2in2] {}
    node at (8.45,-5) [output, name=EqF2out1] {}
    node at (8.45,-6.5) [output, name=EqF2out2] {}
    
    node at (11,-5.75) [block, fill=white, minimum height = 2cm] (phi2) {\Large $\phi^{(2)}_{(\mring{\mb{R}},\mring{\bs{\omega}})}$}
    node at (10.3,-5) [input, name=phi2in1] {}
    node at (10.3,-6.5) [input, name=phi2in2] {}
    node at (11.75,-5) [input, name=phi2out1] {}
    node at (11.75,-6.5) [input, name=phi2out2] {}

    node at (14,-5) [input, name=stage2out1] {}
    node at (14,-6.5) [input, name=stage2out2] {}
;
    \draw[->](camera) -- node {\Large $\mb{R}^\T\mring{\mb{d}}_i$}(EqF2in2);
    \draw[->](EqF2out1) -- node {\Large $\hat{\mb{Q}}$} (phi2in1);
    \draw[->](EqF2out2) -- node {\Large $\hat{\mb{q}}$} (phi2in2);
    \draw[->](phi2out1) -- node {\Large $\hat{\mb{R}}$} (stage2out1);
    \draw[->](phi2out2) -- node {\Large $\hat{\bs{\omega}}$} (stage2out2);

\draw
	node at (1.5,-3) [sum, name=suma2] {\suma}
    node at (12.75,-1.52) {\textbullet}
    node at (1.5,-1.52) {\textbullet}

    ;
    \draw[->] (1.5,-1.5) -- (suma2);
    \draw[->] (12.75,-1.5) |- node[above, xshift=-10.5cm] {\Large $-$} (suma2);
    \draw[->] (suma2) |- node[above, xshift=-1.25cm] {\Large $\mb{u} + \mb{b} - \hat{\mb{b}}$} (EqF2in1);
    
\end{tikzpicture}
    }%
    \caption{Equivariant Filter Cascade}
    \label{fig:equivariant_filter_cascade}
\end{figure}
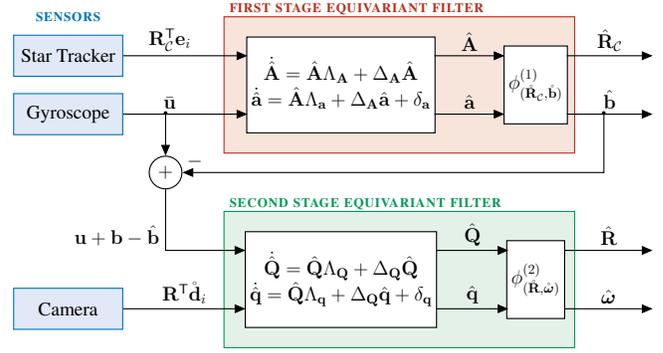

The two derived filters work in a cascaded fashion; consequently, we must study the stability of the equivariant filter cascade to analyze the convergence of all estimated quantities. A diagram of the filter cascade is presented in Fig.\;\ref{fig:equivariant_filter_cascade}. 

Let us analyze the convergence conditions for the first stage filter. If ${\bar{\mb{u}}}$ and ${\hat{\mb{a}}}$ remain bounded for all time, then ${\big(\mring{\mb{A}}_t^{(1)}, {\mb{C}_t}^{(1)}\big)}$ is uniformly completely observable\;\cite{bristeau_design_2010} and the second derivative of the error dynamics is bounded, meaning that ${\bs{\Sigma}^{(1)}(t)}$ is bounded above and below, and the Riccati state equation is well-defined for all time\;\cite{deylon_uniform_observability_2001}. Hence, if ${\bs{\varepsilon}^{(1)}(0)}$ is sufficiently small, these conditions guarantee that ${\bs{\varepsilon}^{(1)} \to \mb{0}}$\;\cite[Theorem 12.1]{krener_convergence_2003}.

The convergence analysis for the second stage filter follows similar steps and was conducted in\;\cite{serrano_eqf_relative_attitude_2025}. Provided that ${\hat{\mb{q}}}$ remains bounded for all time, the pair ${\big(\mring{\mb{A}}_t^{(2)}, {\mb{C}_t}^{(2)}\big)}$ is also uniformly completely observable and the second derivative of the error dynamics is bounded. As such, ${\bs{\Sigma}^{(2)}(t)}$ is bounded above and below, and the Riccati state equation is well-defined for all time. If ${\bs{\varepsilon}^{(2)}(0)}$ is sufficiently small, then ${\bs{\varepsilon}^{(2)} \to \mb{0}}$.

\revised{For each stage ${i \in \{1,2\}}$, we consider the following candidate Lyapunov function}
\begin{equation}
    V^{(i)}\big(\bs{\varepsilon}^{(i)}, t\big) = {\bs{\varepsilon}^{(i)}}^\T {\bs{\Sigma}_{t}^{(i)}}^{-1} \bs{\varepsilon}^{(i)},
\end{equation}
\revised{where $\bs{\Sigma}_{t}^{(i)}\equiv\bs{\Sigma}^{(i)}(t)$ is bounded above and below. The time derivative of $V^{(i)}$ along the trajectories of the linearized error dynamics} \eqref{eq:eqf1_linearized_error_dynamics} and \eqref{eq:eqf2_linearized_error_dynamics} is
\begin{equation}
        \dot{V}^{(i)} = -{\bs{\varepsilon}^{(i)}}^\T \mb{P}_{t}^{(i)} \bs{\varepsilon}^{(i)} \leq -\lambda_{\min}\big\{\mb{P}_{t}^{(i)}\big\} \norm{\bs{\varepsilon}^{(i)}}^{2},
\end{equation}
\revised{where ${\mb{P}_{t}^{(i)}\equiv\mb{C}_{t}^{\top} {\mb{N}^{(i)}_{t}}^{-1} \mb{C}_{t} + {\bs{\Sigma}_{t}^{(i)}}^{-1} {\mb{M}}_{t}^{(i)} {\bs{\Sigma}_{t}^{(i)}}^{-1}}$ is a positive definite matrix for all time. This implies that ${\dot{V}^{(i)} < 0}$ for all ${\bs{\varepsilon}^{(i)} \neq \mb{0}}$,
which shows that both error dynamics are locally uniformly asymptotically stable.
}


However, the expected input in the derivation of the second stage filter is the unbiased angular velocity of the chaser, $\mb{u}$. As such, we take the bias estimate from the first stage filter and subtract it from the gyroscope measurement, ${\bar{\mb{u}}}$, and feed in the second stage filter ${\bar{\mb{u}} - \hat{\mb{b}} = \mb{u} + \mb{b} - \hat{\mb{b}}}$, which converges to $\mb{u}$ as $\hat{\mb{b}}$ converges to $\mb{b}$. We note that $\hat{\mb{b}}$ converging to $\mb{b}$ is equivalent to ${ \mb{b} - \hat{\mb{b}} = \hat{\mb{A}}^\T \bs{\varepsilon}_{\mb{b}} }$ converging to zero, or simply, ${\bs{\varepsilon}_{\mb{b}} \to \mb{0}}$ .
It follows that an accurate estimate of the relative attitude and the target's angular velocity by the second stage filter depends on an accurate estimate of the gyroscope bias by the first stage filter.
Since the second stage filter was developed considering that the input to the system was $\mb{u}$, we must now study the case where true input is ${\bar{\mb{u}} - \hat{\mb{b}}}$. In the following analysis, we drop the superscripts related to the second stage to improve readability. The perturbed error dynamics for the second stage EqF in the local coordinates are
\begin{equation}
\begin{split}
    \dot{\bs{\varepsilon}} &= D\vartheta \cdot D\phi_{\vartheta^{-1}(\bs{\varepsilon})} \big( \Lambda\big(\vartheta^{-1}(\bs{\varepsilon}), (\mring{\mb{u}},\mring{\mb{v}},\mring{\mb{w}})\big) \\
    &\hspace{4mm}-\Lambda\big((\mring{\mb{R}}, \mring{\bs{\omega}}), (\mring{\mb{u}},\mring{\mb{v}},\mring{\mb{w}})\big) - \Gamma_{(\hat{\mb{A}}, \hat{\mb{Q}}, \hat{\mb{q}})}\big(\bs{\varepsilon}^{(1)}\big) \big) \\
    &\hspace{4mm}- D\vartheta \cdot D\phi_{\mb{e}}(\Delta_{\mb{Q}},\delta_{\mb{q}})
\end{split}
\label{eq:perturbed_eqf2_local_coordinates_error_dynamics}
\end{equation}
where ${\Gamma_{(\hat{\mb{A}}, \hat{\mb{Q}}, \hat{\mb{q}})}\big(\bs{\varepsilon}^{(1)}\big) = ((\hat{\mb{Q}}\hat{\mb{A}}^\T \bs{\varepsilon}_{\mb{b}})^{\wedge}, (\hat{\mb{q}})^{\wedge}\hat{\mb{Q}}\hat{\mb{A}}^\T \bs{\varepsilon}_{\mb{b}})}$.
%

Note that the system \eqref{eq:eqf2_local_coordinates_error_dynamics} is the same as the system \eqref{eq:perturbed_eqf2_local_coordinates_error_dynamics} when ${\bs{\varepsilon}^{(1)} = \mb{0}}$.
\revised{However, the estimates ${(\hat{\mb{Q}},\hat{\mb{q}})}$ are affected by the bias. As a consequence, the matrices $\mring{\mb{A}}_{t}^{(2)}$, $\mb{C}_{t}^{(2)}$, and $\bs{\Sigma}_{t}^{(2)}$ are affected, which in turn influence the correction terms ${(\Delta_{\mb{Q}},\Delta_{\mb{q}})}$.}
\revised{Nevertheless, as long as ${\hat{\mb{q}}}$ remains bounded for all time with the input $\bar{\mb{u}}-\hat{\mb{b}}$, then the effect on the matrices and correction terms do not void the conditions for the local uniform asymptotic stability of the second stage filter.}
Therefore, since both \eqref{eq:eqf1_local_coordinates_error_dynamics} and \eqref{eq:eqf2_local_coordinates_error_dynamics} are locally uniformly asymptotically stable, by\;\cite[Theorem 3.1]{vidyasagar_decomposition_1980}, we conclude that the cascaded system composed of \eqref{eq:eqf1_local_coordinates_error_dynamics} and \eqref{eq:perturbed_eqf2_local_coordinates_error_dynamics} is also locally uniformly asymptotically stable.

\section{Simulation Results}\label{sec:simulation_results}

To assess the performance of the equivariant filter cascade, we carry out numerical simulations in MATLAB\textsuperscript{\textregistered}. We conduct 1000 Monte Carlo simulations to evaluate the filters' performance and analyze in detail the results of a representative run. We also study the effect of gyroscope bias estimation and the impact of low measurement rates on the filter's performance. 

\subsection{Simulation Setup}

In each simulation run, the chaser and target are initialized with random attitudes, meaning that the initial relative attitude is also random. To recover the system of interest, we set the virtual inputs that were introduced to prove equivariance to zero, ${\mb{v} = \mb{w} = \mb{0}}$, and assume a random constant angular velocity $\bs{\omega}_\mc{T}$ for the target. Though not required in our approach, we set the chaser angular velocity, $\mb{u}$, to also be constant and randomly generated. The constant gyroscope bias is also randomly defined.
We assume noisy measurements of the gyroscope data, with additive white noise that follows the distribution $\mathpzc{N}(\mb{0}, 0.01^2 \mb{I})$. The star tracker and direction vector measurements are also assumed to include noise. This is simulated by rotating the noiseless measurements about a random axis, uniformly distributed over the 2-sphere, by an angle sampled from a Gaussian distribution $\mathpzc{N}(0, 0.01^2)$.

The initial filter estimates in the group are set to ${(\hat{\mb{A}},\hat{\mb{a}}) = (\mb{I}, \mb{0}})$ and ${(\hat{\mb{Q}},\hat{\mb{q}}) = (\mb{I}, \mb{0}})$, with gain matrices ${\mb{M}^{(1)}_t = \mb{M}^{(2)}_t = \mb{I}}$ and ${\mb{N}^{(1)}_t = \mb{N}^{(2)}_t = 0.1\,\mb{I}}$. The initial value of the Riccati states are ${\bs{\Sigma}^{(1)}_0 = \bs{\Sigma}^{(2)}_0 = \mb{I}}$.
The filters are implemented in a prediction-update sequence, with the prediction step of each filter running at \SI{100}{\hertz}, which is the assumed gyroscope measurement frequency, and the update step frequency depending on the corresponding sensor measurement rate, as will be explained. The system is simulated for a total of \SI{15}{\second} in each run.

\subsection{First Stage Equivariant Filter}

\begin{figure}
    \centering
    \begin{minipage}{\columnwidth}
            \centering
            \begin{subfigure}[t]{\textwidth}
                \centering
                \includegraphics[width=\textwidth]{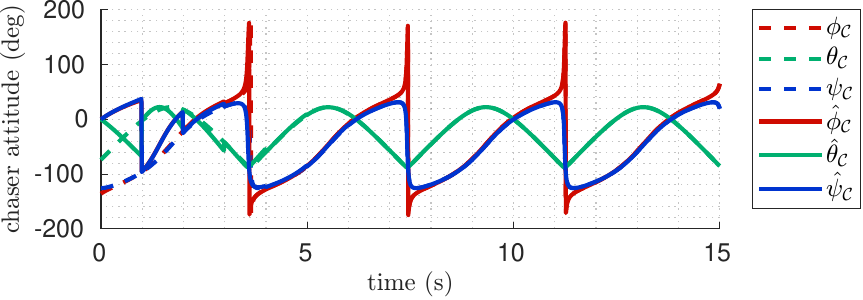} 
                \caption{Chaser's attitude, $\mb{R}_{\mc{C}}$.}
                \label{fig:sim_attitude_chaser}
            \end{subfigure}
            \begin{subfigure}[t]{\textwidth}
                \centering
                \includegraphics[width=\textwidth]{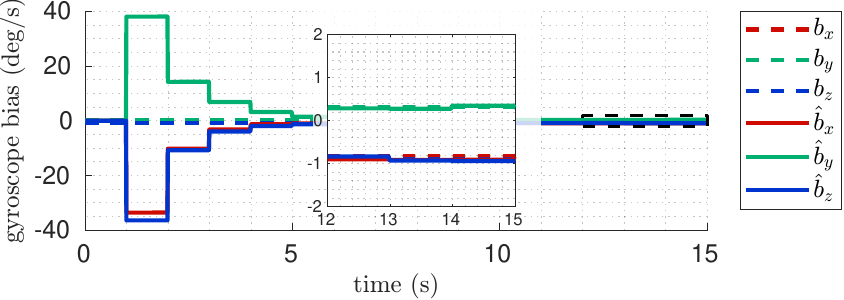} 
                \caption{Gyroscope bias, $\mb{b}$.}
                \label{fig:sim_bias}
            \end{subfigure}
    \end{minipage}
    \caption{First Stage EqF: True (dashed lines) and estimated (solid lines) chaser's attitude and gyroscope bias.}
    \label{fig:sim_first_filter}
\end{figure}

For the EqF in the first stage of the cascade, we assume that the star tracker measurements are available at a rate of \SI{1}{\hertz}. Hence, the update step of this filter is calculated at this frequency. If we were to do a regular update step at such a low frequency, the filter would not converge. Therefore, with each star tracker measurement, the update step is iterated several times (e.g., 20 times, for the results presented here), by following the iterative strategy explained in\;\cite{serrano_eqf_relative_attitude_2025}.
In Fig.\;\ref{fig:sim_first_filter}, we present the true values and the estimates of the chaser's attitude and the gyroscope bias of a representative simulation. For visualization purposes, the chaser's attitude is reported in Euler angles, with $\phi_\mc{C}, \theta_\mc{C}, \psi_\mc{C}$ being the roll, pitch, and yaw, respectively.

In the time interval between {10} and \SI{15}{\second}, after the filter estimates converge, the mean of the chaser's attitude error norm is \num{1.8e-3}, which corresponds to a mean error norm of \SI{0.069}{\degree}, \SI{0.046}{\degree}, and \SI{0.069}{\degree}, for roll, pitch, and yaw, respectively.
The mean of the gyroscope bias error norm is \SI{0.082}{\degree\per\second}, that is, a relative error of \SI{6.58}{\percent} with respect to the bias norm, which was approximately \SI{1.25}{\degree\per\second} in this simulation. 

\subsection{Second Stage Equivariant Filter}
\begin{figure}
    \centering
    \begin{minipage}{\columnwidth}
            \centering
            \begin{subfigure}[t]{\textwidth}
                \centering
                \includegraphics[width=\textwidth]{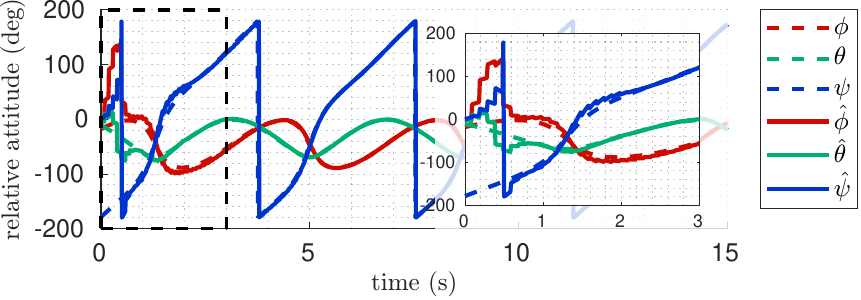} 
                \caption{Relative attitude, $\mb{R}$.}
                \label{fig:sim_attitude_relative}
            \end{subfigure}
            \begin{subfigure}[t]{\textwidth}
                \centering
                \includegraphics[width=\textwidth]{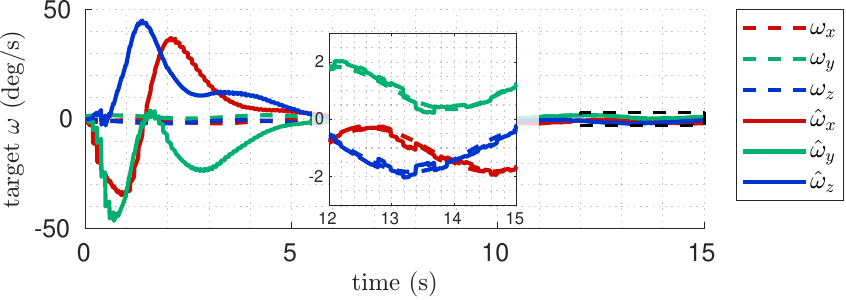} 
                \caption{Target's angular velocity, $\bs{\omega}$.}
                \label{fig:sim_angular_velocity_target}
            \end{subfigure}
    \end{minipage}
    \caption{Second Stage EqF: true (dashed lines) and estimated (solid lines) relative attitude and target's angular velocity.}
    \label{fig:sim_second_filter}
\end{figure}

For the second stage EqF, we assume that the direction vectors are measured at a frequency of \SI{10}{\hertz} and, as such, this is the frequency of the update step. As before, with each measurement, the update step is iterated 20 times, even though a \SI{10}{\hertz} measurement rate is sufficient for convergence. Despite not being strictly necessary, employing the iterative strategy yields better results.
In Fig.\;\ref{fig:sim_second_filter}, we show the true values and the estimates of the relative attitude and the target's angular velocity. Again, to aid in visualization, the relative attitude is reported in Euler angles, with $\phi$, $\theta$, $\psi$ being the roll, pitch, and yaw, respectively.

In the time interval between {10} and \SI{15}{\second}, the mean of the relative attitude error norm is \num{6.3e-3}, which corresponds to a mean error norm of \SI{0.207}{\degree}, \SI{0.184}{\degree}, and \SI{0.172}{\degree}, for roll, pitch, and yaw, respectively.
The mean of the target's angular velocity error norm is \SI{0.20}{\degree\per\second}, that is, a relative error of \SI{9.69}{\percent} with respect to the target's angular velocity norm, which was approximately \SI{2.09}{\degree\per\second} in this simulation. It should be noted that this value has the same order of magnitude as the gyroscope bias.
%

\subsection{Monte Carlo Results}

Tables\;\ref{tab:chaser_attitude_performance_metrics} to\;\ref{tab:angular_velocity_performance_metrics} summarize the performance metrics obtained from the 1000 Monte Carlo simulation runs for both stages of the filter cascade. For the chaser's attitude and the relative attitude, we present the mean and the minimum error norms, in the time interval between \num{10} and \SI{15}{\second}, as well the time taken to reach an error norm below \SI{1}{\degree}, for roll, pitch, and yaw. For the gyroscope bias and the target's angular velocity, we present the mean and minimum absolute error norms, along with the relative error norm, with respect to the true bias and angular velocity, respectively, also in the time interval between \num{10} and \SI{15}{\second}. 
Both the first and second stage filters are compared against a scenario where the measurements occur at \SI{100}{\hertz}, without the need to iterate the update step. This allows us to assess the effect of the low measurement rate and the effectiveness of the iterative strategy employed. The second stage filter is also compared against a scenario with a biased input, i.e., without the bias estimation and removal performed by the first stage filter. This allows us to evaluate the effect of the gyroscope bias on the target's angular velocity estimation.
These two cases will be discussed next.

{
\sisetup{
  mode = text,                  
  detect-family = false,        
  detect-weight = true,         
  per-mode = symbol,            
  exponent-product = \cdot,     
  separate-uncertainty = true,  
  round-mode=places,
  round-precision=2,
}
\begin{table}
    \centering
    \caption{Chaser's Attitude Estimation Metrics}
    \resizebox{\linewidth}{!}{%
    \begin{tabular}{
        l  
        l  
        S[table-format=1.3]  
        S[table-format=1.3, round-precision=3]
        S[scientific-notation=true, table-format=1.2e1]  
        }
        \toprule
        \multicolumn{2}{c}{} &
        \multicolumn{1}{c}{\shortstack{Time to\\1\si{\degree} error (\si{\second})}} &
        \multicolumn{1}{c}{\shortstack{Mean error (\si{\degree})\\in [10, 15]\,\si{\second}}} &
        \multicolumn{1}{c}{\shortstack{Minimum\\error (\si{\degree})}} \\
        \midrule
        
        \multirow{3}{*}{Chaser's Attitude} 
          & Roll $\phi_\mc{C}$      & 1.197960 & 0.243463 & 0.001994 \\
          & Pitch $\theta_\mc{C}$   & 1.004590 & 0.038026 & 0.000108 \\
          & Yaw $\psi_\mc{C}$       & 1.194780 & 0.243747 & 0.002046 \\
        \midrule
        \multirow{3}{*}{\shortstack{Chaser's Attitude\\(100\,Hz)}} 
          & Roll $\phi_\mc{C}$      & 0.473850 & 0.067878 & 0.000190 \\
          & Pitch $\theta_\mc{C}$   & 0.341170 & 0.013249 & 0.000043 \\
          & Yaw $\psi_\mc{C}$       & 0.475960 & 0.067784 & 0.000161 \\
        \bottomrule
    \end{tabular}
    }
    \label{tab:chaser_attitude_performance_metrics}
\end{table}
}

        

{
\sisetup{
  mode = text,                  
  detect-family = false,        
  detect-weight = true,         
  per-mode = symbol,            
  exponent-product = \cdot,     
  separate-uncertainty = true,  
  round-mode=places,
  round-precision=2,
}
\begin{table}
    \centering
    \caption{Gyroscope Bias Estimation Metrics}
    \resizebox{\linewidth}{!}{%
    \begin{tabular}{c
                    S[table-format=1.3,round-precision=3]
                    S[table-format=1.3]
                    S[table-format=1.3,round-precision=3]
                    S[table-format=1.3]
                    }
        \toprule
        {} & 
        {\shortstack{Mean error ($\si{\degree\per\second}$)\\in [10, 15]\,s}} & 
        {\shortstack{Mean error ($\si{\percent}$)\\in [10, 15]\,s}} & 
        {\shortstack{Minimum\\error ($\si{\degree\per\second}$)}}  & 
        {\shortstack{Minimum\\error ($\si{\percent}$)}} \\
        \midrule
        Gyroscope Bias                               & 0.066733 & 6.020076 & 0.031436 & 1.620916 \\
        \midrule
        \makecell[c]{Gyroscope Bias\\(100\,Hz)}      & 0.05911 & 5.290073 & 0.015082 & 0.774734 \\
        \bottomrule
    \end{tabular}
    }
\label{tab:gyroscope_bias_performance_metrics}
\end{table}
}

{
\sisetup{
  mode = text,                  
  detect-family = false,        
  detect-weight = true,         
  per-mode = symbol,            
  exponent-product = \cdot,     
  separate-uncertainty = true,  
  round-mode=places,
  round-precision=2,
}
\begin{table}
    \centering
    \caption{Relative Attitude Estimation Metrics}
    \resizebox{\linewidth}{!}{%
    \begin{tabular}{
        l  
        l  
        S[table-format=1.3]
        S[table-format=1.3,round-precision=3]  
        S[scientific-notation=true, table-format=1.2e1]  
        }
        \toprule
        \multicolumn{2}{c}{} &
        \multicolumn{1}{c}{\shortstack{Time to\\1\si{\degree} error (\si{\second})}} &
        \multicolumn{1}{c}{\shortstack{Mean error (\si{\degree})\\in [10, 15]\,\si{\second}}} &
        \multicolumn{1}{c}{\shortstack{Minimum\\error (\si{\degree})}} \\
        \midrule
        \multirow{3}{*}{\shortstack{Relative Attitude\\(with biased input)}}
          & Roll $\phi$     & 0.7014 & 0.1812 & 0.000924 \\
          & Pitch $\theta$  & 0.5945 & 0.1271 & 0.000744 \\
          & Yaw $\psi$      & 0.5737 & 0.1616 & 0.000818 \\
        \midrule
        \multirow{3}{*}{\shortstack{Relative Attitude\\(with unbiased input)}}
          & Roll $\phi$     & 0.7010 & 0.2042 & 0.001202 \\
          & Pitch $\theta$  & 0.5886 & 0.1284 & 0.000762 \\
          & Yaw $\psi$      & 0.5799 & 0.1848 & 0.000822 \\
        \midrule
        \multirow{3}{*}{\shortstack{Relative Attitude\\(with unbiased, 100\,Hz)}}
          & Roll $\phi$     & 0.7211 & 0.0695 & 0.000239 \\
          & Pitch $\theta$  & 0.4964 & 0.0444 & 0.000150 \\
          & Yaw $\psi$      & 0.5863 & 0.0627 & 0.000219 \\
        \bottomrule
    \end{tabular}
    }
    \label{tab:relative_attitude_performance_metrics}
\end{table}
}

{
\sisetup{
  mode = text,                  
  detect-family = false,        
  detect-weight = true,         
  per-mode = symbol,            
  exponent-product = \cdot,     
  separate-uncertainty = true,  
  round-mode=places,
  round-precision=2,
}
\begin{table}
    \centering
    \caption{Target's Angular Velocity Estimation Metrics}
    \resizebox{\linewidth}{!}{%
    \begin{tabular}{l
                    S[table-format=1.2]
                    S[table-format=2.2]
                    S[table-format=1.3,round-precision=3]
                    S[table-format=2.2]
                    }
        \toprule
        {} & 
        {\shortstack{Mean error ($\si{\degree\per\second}$)\\in [10, 15]\,s}} & 
        {\shortstack{Mean error ($\si{\percent}$)\\in [10, 15]\,s}} & 
        {\shortstack{Minimum\\error ($\si{\degree\per\second}$)}} & 
        {\shortstack{Minimum\\error ($\si{\percent}$)}}  \\
        \midrule
        \makecell[c]{Target's Ang. Vel.\\(with biased input)} & 1.192081 & 81.776306 & 0.816856 & 32.044899 \\
        \midrule
        \makecell[c]{Target's Ang. Vel.\\(with unbiased input)} & 0.187535 & 12.834685 & 0.040585 & 1.592046 \\
        \midrule
        \makecell[c]{Target's Ang. Vel.\\(with unbiased\\input, 100\,Hz)} & 0.071951 & 4.930969 & 0.008776 & 0.344105\\
        \bottomrule
    \end{tabular}
    }
\label{tab:angular_velocity_performance_metrics}
\end{table}
}


\subsubsection{Effect of Gyroscope Bias Estimation}

Let us first analyze the effect of gyroscope bias estimation and its subsequent removal from the measurements on the relative attitude and the target's angular velocity estimation.

\begin{figure}
    \centering
    \includegraphics[width=\columnwidth]{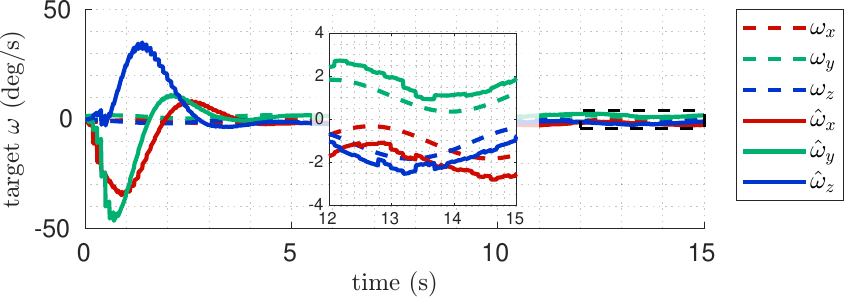}
    \caption{Target's angular velocity, $\bs{\omega}$, with biased filter input.}
    \label{fig:angular_velocity_target_biased}
\end{figure}

From the results presented in Table\;\ref{tab:relative_attitude_performance_metrics}, one concludes that the presence of bias does not significantly influence the relative attitude estimation. Note that the system's measurements, defined in \eqref{eq:system_output_y_2}, are directly related to the relative attitude, thus mitigating the unbiased input's effect. However, in Table\;\ref{tab:angular_velocity_performance_metrics}, we can see that, without the unbiased gyroscope measurements, in the time interval between \num{10} and \SI{15}{\second}, over all runs, the target's angular velocity mean error norm goes from \SI{0.19}{\degree\per\second} (\SI{12.83}{\percent}) up to \SI{1.19}{\degree\per\second} (\SI{81.78}{\percent}) and the minimum error norm goes from \SI{0.041}{\degree\per\second} (\SI{1.59}{\percent}) up to \SI{0.817}{\degree\per\second} (\SI{32.04}{\percent}). This effect can be seen in Fig.\;\ref{fig:angular_velocity_target_biased}. Using the same scenario as the representative run previously analyzed, without the bias removal, the estimated target's angular velocity has an offset with respect to the true value.
As stated in Section\;\ref{sec:introduction}, the angular velocity is not fully observable in the presence of the gyroscope bias. The fact that the target's angular velocity and the gyroscope bias have the same order of magnitude significantly contributes to the discrepancy in performance. In such cases, the unbiased input to the second stage filter is necessary for a more accurate estimation. 
However, if the difference between the target's angular velocity and the bias were larger, the bias would be less significant and the negative effect would be mitigated.

\subsubsection{Effect of Measurement Rate}

Let us now analyze the effect that the low measurement rates have on the filtering method proposed. Besides the case already presented, with the star tracker measurements at a rate of \SI{1}{\hertz} and the vector measurements at \SI{10}{\hertz}, we will also consider a scenario where both the star tracker and vector measurements occur at \SI{100}{\hertz}. Being the same rate as the gyroscope measurements, this means that the prediction and update steps are calculated sequentially, without the need to iterate the update step after each measurement. 

From the results presented in Tables\;\ref{tab:chaser_attitude_performance_metrics} to\;\ref{tab:angular_velocity_performance_metrics}, one concludes that a faster measurement rate reduces the estimation error of all quantities. It particularly improves the estimation of the target's angular velocity, going from a \SI{12.83}{\percent} mean relative error norm to \SI{4.93}{\percent} and from a \SI{1.59}{\percent} minimum relative error norm to \SI{0.34}{\percent}, averaged over all the simulation runs. The estimates of the chaser's attitude and the gyroscope bias also benefit from the increased measurement rate, though to a lesser extent.

\begin{figure}[ht]
    \centering
    \begin{minipage}{\columnwidth}
            \centering
            \begin{subfigure}[t]{\textwidth}
                \centering
                \includegraphics[width=\textwidth]{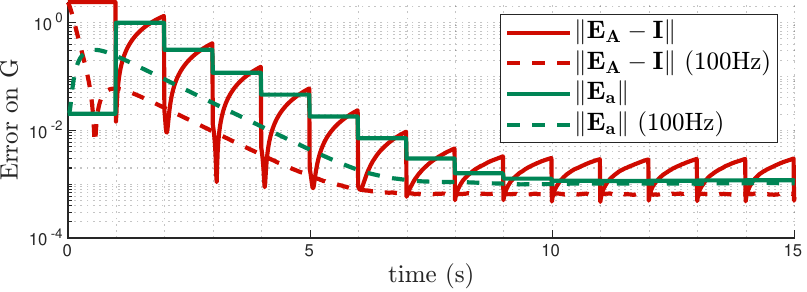} 
                \caption{First Stage EqF: group error norm.}
                \label{fig:error_group_log_montecarlo_stage1}
            \end{subfigure}
            \begin{subfigure}[t]{\textwidth}
                \centering
                \includegraphics[width=\textwidth]{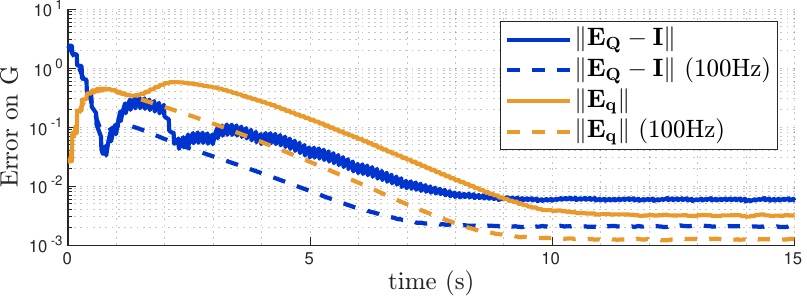} 
                \caption{Second Stage EqF: group error norm.}
                \label{fig:error_group_log_montecarlo_stage2}
            \end{subfigure}
    \end{minipage}
    \caption{Norm of the errors on the group $\mr{G}$, in log-scale, for different measurement rates. Solid lines represent the scenario with star tracker measurements at \SI{1}{\hertz} and vector measurements at \SI{10}{\hertz}, while dashed lines represent the scenario with measurements at \SI{100}{\hertz}.}
    \label{fig:error_comparison_rate}
\end{figure}

Fig.\;\ref{fig:error_comparison_rate} displays, in log-scale, the norm of the error of the estimated quantities in $\mr{G}$ in the two scenarios, averaged over the Monte Carlo simulations. Note that $\mb{E}_{\mb{A}}$ and $\mb{E}_{\mb{Q}}$ appear with the identity subtracted so that the error norm converges to zero. We can observe that, though the filtering strategy converges in both scenarios, it is more accurate and shows a smoother and slightly faster convergence with a faster measurement rate (dashed lines). It is important to emphasize that, in the first scenario, this is only possible because the negative impact of the low measurement rate is compensated by iterating the update step with each measurement; otherwise, the filter cascade would not converge. The second case, though not practically attainable due to the sensors' limitations, demonstrates the performance that the designed equivariant filter cascade could theoretically have.


\subsection{Discussion} 

Through the above analysis, it is clear that unbiased gyroscope measurements, used as input to the second stage filter, are important for an accurate estimation of the target's angular velocity. The unbiased measurements, obtained via the first stage filter, become even more essential when the target's angular velocity and gyroscope bias are similar in magnitude. 
In addition, we noted that the low sensor measurement rate also impacts the performance and an iteration-based strategy was employed to lessen the negative effect. In our previous work\;\cite{serrano_eqf_relative_attitude_2025}, we used an event camera to obtain the vector measurements at a much faster rate (\SI{1}{\kilo\hertz}), thus improving the filter's performance. Other studies have demonstrated the potential of event cameras for star tracking, reporting measurement frequencies between \SI{500}{\hertz} and \SI{1}{\kilo\hertz}\;\cite{ng_asynchronous_2023,reed_event_star_tracking_2025}. Usage of these sensors may thus allow for further performance improvements of the proposed method. 


\addtolength{\textheight}{-11cm}   
\section{Conclusions} \label{sec:conclusions}

In this work, we derived and implemented an equivariant filter cascade. The first stage filter estimates the chaser's attitude and gyroscope bias, whilst the second stage filter estimates the relative attitude between the chaser and the target, as well as the target's angular velocity. 
We evaluated a scenario in which the target spacecraft exhibits a constant angular velocity and studied the effect of low measurement rates on the multi-stage filter performance, using simulated data. Our approach is able to correctly estimate the desired system states and input measurement bias.



\bibliographystyle{IEEEtran}
\bibliography{mybibfile}

\end{document}